# Personal Identification Using Ultrawideband Radar Measurement of Walking and Sitting Motions and a Convolutional Neural Network[†]

### Takuya Sakamoto

*Abstract*— **This study proposes a personal identification technique that applies machine learning with a two-layered convolutional neural network to spectrogram images obtained from radar echoes of a target person in motion. The walking and sitting motions of six participants were measured using an ultrawideband radar system. Time-frequency analysis was applied to the radar signal to generate spectrogram images containing the micro-Doppler components associated with limb movements. A convolutional neural network was trained using the spectrogram images with personal labels to achieve radar-based personal identification. The personal identification accuracies were evaluated experimentally to demonstrate the effectiveness of the proposed technique.**

## I. Introduction

NEXT-GENERATION sensor technology applications have attracted increased attention in recent years for realizing an ultra-smart society or Society 5.0, one of the strategic goals of the Japanese Cabinet Office. Society 5.0 is expected to be capable of "providing the needed goods and services to the people who need them, when they need them, and as much as they need" [1]. Realizing Society 5.0 will require sensor technology to monitor the positions, movements, and biological signals of people as well as automatic personal identification technology. Currently, facial recognition technology based on camera images has attracted much attention [2] [3], and the use of security cameras in outdoor and public facilities is increasing rapidly. Further, personal identification using smart-speaker-based voiceprint matching

† Translated from IEICE Transactions Japanese Edition, vol. J103-C, no. 07, pp. 321-330, July 2020 (in Japanese).

This study was supported in part by JSPS KAKENHI 19H02155, 15K18077, and 15KK0243; JST PRESTO JPMJPR1873; and JST COI JPMJCE1307.

T. Sakamoto is with Department of Electrical Engineering, Graduate School of Engineering, Kyoto University, Kyoto 615-8510, Japan (e-mail: sakamoto.takuya.8n@kyoto-u.ac.jp).

T. Sakamoto is also with the PRESTO, Japan Science and Technology Agency, 4-1-8 Honcho, Kawaguchi, Saitama 332-0012, Japan.

has become popular in indoor environments. These devices have become very popular because of their low cost and high accuracy; however, they raise privacy concerns that cannot be ignored. For example, information recorded from the cameras and microphones of these devices has often been leaked, and therefore, the use of such devices has produced much social anxiety.

In this light, personal identification sensor technology must be developed in consideration of user privacy. For example, biometric authentication methods like fingerprint authentication [4], [5], iris identification [6], vein identification [7-10], and eye movement identification [11] do not require entire face or body images or voice recordings. Studies have also performed personal identification by using biometric information such as electrocardiograms (ECGs) [12-14] and electroencephalograms [15], [16]. For example, Koike-Akino et al. [17] and Louis et al. [18] extracted characteristic quantities from ECG data to perform biometric authentication. Further, Odinaka et al. [19] conducted a literature review of various ECG-based personal identification techniques. Many of these techniques, however, are cumbersome to use; for example, they require electrodes to be attached to the body surface or the sensor to be brought close to the body.

Radio wave sensing using radar has attracted research attention as a simple noncontact method. For example, Diederichs et al. [20] performed personal identification through millimeter-wave radar measurements of reflected waves from the subcutaneous tissue of the palm. Although their approach achieved noncontact personal identification, users had to bring their hand close to the antenna. Dwelly and Adams [21] estimated the posture of subjects by performing radar measurements and processing the received signals in a neural network, but they did not perform personal identification. Rissacher and Galy [22] measured heartbeats with a 2.4-GHz radar and investigated its potential applicability to biometric authentication. Similarly, Shi et al. [23] achieved a personal identification success rate of 91.3% through radar measurements of the heartbeats of four subjects. Rahman et al. [24] achieved a personal identification success rate of 95% through 2.4-GHz radar measurements of the breathing of six subjects. Radar-based measurements therefore show some promise for the personal identification of biological signals such as heartbeats and breathing; however, they currently find limited applications because they require subjects to be still or in a resting state. Furthermore, careful examination is required prior to introduction, and the reproducibility has been reported to decrease with changes in physical condition.

By contrast, personal identification based on micro-Doppler measurements of limb motions, such as those during walking, shows more promise. Garreau et al. [25] reported a personal identification success rate of 100% with ultrasonic sonar measurements of 13 subjects who were walking on a treadmill 2.5 m away from a sensor. The speed and position of the subjects were controlled as they were walking on the treadmill, thereby potentially improving the accuracy. Kalgaonkar and Raj [26] reported a personal identification success rate of 91.7% with mixed Gaussian models through 20 repeated ultrasonic sonar measurements of the walking motion of 30 subjects. Unfortunately, ultrasonic waves with high resolution undergo large attenuation in air, and therefore, such sonar-based human measurements are only possible over relatively short distances.

To improve radar-based measurements, Cao et al. [27] used a 24-GHz K-band radar with 250 MHz bandwidth to perform 50 repeated measurements of the walking motion of 24 subjects and achieved personal identification from spectrogram images using AlexNet, a type of five-layer convolutional neural network (CNN). They achieved personal identification success rates of 90.9% and 68.9% with six and 20 subjects, respectively. Vandersmissen et al. [28] used a 77-GHz-band and 1.5-GHz-bandwidth radar to measure five subjects freely walking indoors and achieved a personal identification success rate of 78.4% by inputting the spectrogram images into a four-layer CNN. Tahmoush and Silvious [29] achieved a personal identification success rate of 80% by measuring the walking motion of eight subjects using radar, manually extracting features using micro-Doppler measurements, and using the k-nearest neighbors (KNN) algorithm. Overall, these conventional studies performed personal identification by measuring walking motions.

By contrast, Yang et al. [30] conducted 4.3-GHz ultrawideband radar measurements of 15 subjects for six types of movement including walking, shadowboxing while walking, crawling forward, tip-toeing, jumping, and running. They achieved personal identification success rates of 72.6%, 94.4%, and 95.2% for crawling, walking, and running, respectively, by inputting spectrogram images into a multilayer CNN. Seyfioğlu et al. [31] measured 12 types of movement, including walking, in 11 subjects with a 4-GHz continuous wave (CW) radar. Their study of movement identification, and not personal identification, achieved the following success rates: with a support vector machine that used 50-dimensional characteristic quantities, 76.9%; with an auto-encoder, 84.1%; with a convolutional neural network, 90.1%; and with a convolutional auto-encoder, 94.2%.

The present study performs micro-Doppler-based personal identification of the series of motions from walking to sitting. To the best of the author's knowledge, no study has performed personal identification using radar measurements of sitting motions. Radar waves are irradiated toward a fixed chair to identify users who approach and sit in this chair. This method could be applied to users who sit at computer terminals with highly confidential electronic files or to detect suspicious behaviors in secure facilities. In the experiments, an ultrawideband radar was used to measure the walking and sitting motions of six subjects. A time-frequency distribution was obtained from the received signal, and personal identification was conducted using a CNN. The performance was quantitatively evaluated in terms of the personal identification accuracy rate using data with both walking and sitting motions that lasted ~8 s. Note that some of the preliminary investigations for this study have been orally presented in [32].

## II. RADAR-BASED HUMAN BODY MEASUREMENTS AND MICRO-DOPPLER

A Doppler shift is observed in echoes from a person in motion. Each part of the human body generally shows a different type of motion; therefore, the received signal is a superposition of multiple echoes corresponding to each body part, and each Doppler shift amount varies with time. The average movement speed of the entire human body corresponds to the conventional Doppler shift of radar echoes, whereas micro-Doppler is a collection of Doppler components caused by the moving components of the target, such as the limbs of the human body, and is a quasi-periodic function of time.

The signal measured with ultrawideband radar is expressed as a function $s_0(t, r)$ of time $t$ and range $r$. Here, the range $r$ is expressed with a delay time $\tau$ as $r = c\tau/2$. Here, $c$ is the speed of the radio wave. The pulse-compressed signals are directly A/D converted without down-converting; therefore, $s_0(t, r)$ is a real value. First, $s_0(t, r)$ is converted to an analytic signal, $s(t, r)$, by the following equation:

$$s(t, r) = \frac{1}{2\pi} \int_{-\infty}^{\infty} U(k) e^{jkr} \int_{-\infty}^{\infty} s_0(t, r') e^{-jkr'} dr' \, dk \quad (1)$$

Here, $U(k)$ is a unit step function. Next, a time-frequency distribution (spectrogram) $S(t, v_d)$ is obtained using a short-time Fourier transform (STFT).

$$S(t, v_d) = \int_{r_1}^{r_2} \left| \int_{-\infty}^{\infty} w(t') s(t'-t, r) e^{-j2kv_d t'} dt' \right|^2 dr \quad (2)$$

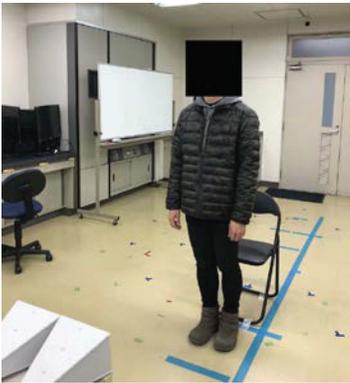

Fig. 1. Ultrawideband radar-based measurement scene of subject.

Here, $k$ is the wave number, $v_{\mathrm{d}}$ is the Doppler velocity, and the window function $w(\cdot)$ is set as a Hann window with a time width of $t_0 = 0.16$ s as shown in the following equation:

$$w(t) = \begin{cases} \frac{1}{2}\left(1 + \cos\left(\dfrac{2\pi t}{t_0}\right)\right) & |t| \leq t_0/2 \\ 0 & |t| > t_0/2 \end{cases} \tag{3}$$

Furthermore, the human body is assumed to exist in the range $r_1 \leq r \leq r_2$. In addition to reflected waves from the human body, the received signal includes clutter, which are unwanted components such as direct waves between transmitting and receiving antennas or reflected waves from static objects in the room. Therefore, only spectrograms within the range $r_1 \leq r \leq r_2$ are calculated. Further, clutter is removed by subtracting a signal measured beforehand when no people are present from the signal that includes the reflected wave from the human body. Some methods can remove the DC component $v_{\mathrm{d}} = 0$; however, this can unintentionally remove the human body echo under low-speed or static conditions, and therefore, it is not used in this study. In the following section, $S(t, v_{\mathrm{d}})$ is saved as an image and used for CNN learning and performance evaluation.

## III. Experimental Setup and Neural Networks

### A. Radar-based Human Motion Measurement

The PulsON® 400 (Time Domain Co., Huntsville, AL, USA) ultrawideband radar with a central frequency of 4.2 GHz and a bandwidth of 2.2 GHz was used for measurements. This radar transmitted a wideband signal that was phase-modulated by a binary pseudo-noise sequence (M sequence) from a transmitting antenna, which was then re-ceived at a separate receiving antenna. The signal obtained by the receiving antenna was then pulse-compressed and stored as data after A/D conversion. A bandwidth of 2.2 GHz corresponds to a range resolution of 6.8 cm; therefore, most unnecessary waves and static clutter could be suppressed while maintaining the desired wave from the human body according to distance. However, some static clutter was superimposed on the desired signal as a DC component after time-frequency analysis, which caused interference. Two wideband double-ridged horn antennas were used for transmitting and receiving signals; they were installed near each other to approximate a monostatic radar.

Six subjects (A, B, C, D, E, and F) walked and sat continuously in a room. First, each subject stood at a distance of ~7.0 m from the transmitting and receiving antennas and began walking toward the antenna in response to the signal to start the measurements. Then, the subject sat after moving to the front of a pipe chair installed at a distance of ~3.5 m from the antenna. As the subject performed this series of motions, radar measurements were performed. Table 1 lists the measurement specifications and Fig. 1 shows a photograph of the measurement scene.

TABLE I
RADAR SYSTEM SPECIFICATIONS

| Parameter | Value |
| --- | --- |
| Antenna | Double-ridged horn |
| Center frequency | 4.2 GHz |
| 10-dB bandwidth | 2.2 GHz |
| Range sampling interval | 9.12 mm |
| Time sampling interval | 5 ms |
| Beamwidth (E-plane) | 32° |
| Beamwidth (H-plane) | 34° |
| Modulation | Binary phase |
| Spreading sequence | M-sequence |

TABLE II
SPECIFICATIONS OF NEURAL NETWORK TO BE USED

| | | |
| --- | --- | --- |
| Input layer size | | $32 \times 100$ |
| Convolutional layer 1 | # of filters | 10 |
| | Filter size | $10 \times 10$ |
| | Activation | ReLU |
| | Max pooling size | $2 \times 2$ |
| | Output size | $10 \times 22 \times 45$ |
| Convolutional layer 2 | # of filters | 10 |
| | Filter size | $10 \times 10 \times 10$ |
| | Activation | ReLU |
| | Max pooling size | $2 \times 2$ |
| | Output size | $10 \times 12 \times 17$ |
| Fully-connected layer | | $2,040 \times 6$ |
| Output layer size | | 6 |

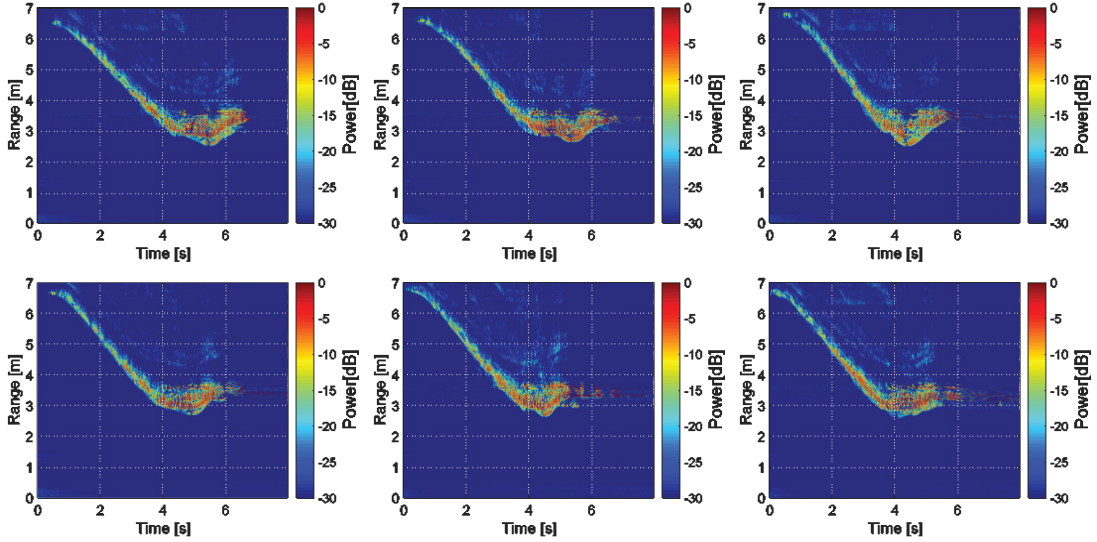

Fig. 2. Intensity distribution of the received signals at each time and range. Subjects A, B, C, D, E, and F, in order from the top left.

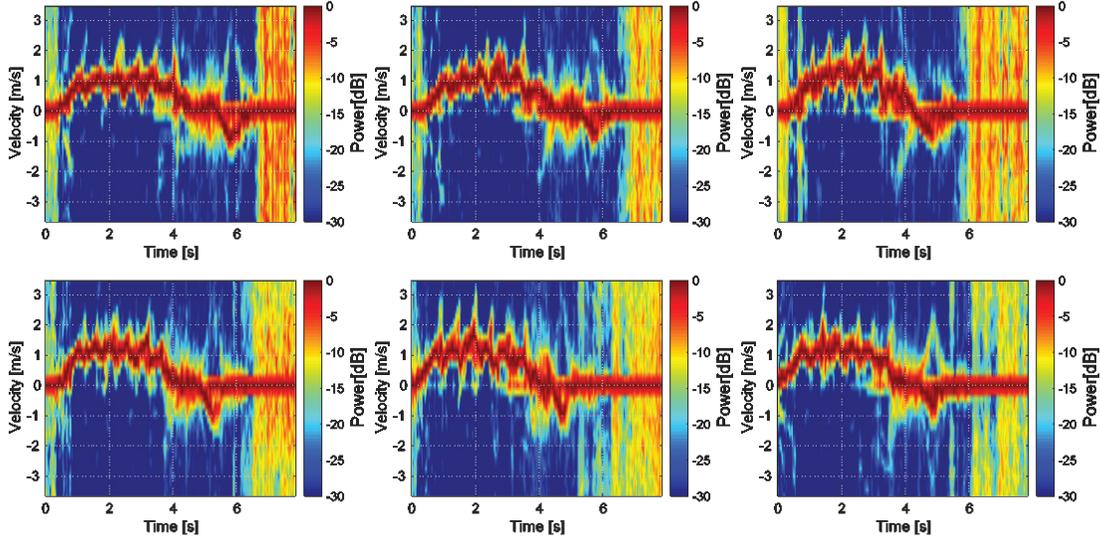

Fig. 3. Time-frequency distribution of received signals. Subjects A, B, C, D, E, and F, in order from the top left.

Each measurement lasted ~8 s, and the measurement was repeated 100 times for each subject. Fig. 2 shows an example of the received signal power $|s(t, r)|^2$ of the six subjects as a function of time $t$ and distance (range) $r$. This figure is normalized to the maximum value. All subjects arrived at the front of the chair ~4 s after measurements and sat down after ~6 s. Fig. 3 shows spectrograms $S(t, v_d)$ obtained using STFT. Here, the direction toward the antennas is set as the positive velocity. Spike-like time-variable components can be seen above and below the representative velocity components. Micro-Doppler components are generated from the limb motions during walking. Furthermore, a negative velocity component can be seen when sitting down; it corresponds to the head and upper body area moving away from the antenna.

### B. Personal Identification Using a Neural Network

Table 2 lists the specifications of the CNN used in this study and Fig. 4 shows the configuration. The input data is a $32 \times 100$-pixel 8-bit grayscale image. A convolution was conducted between the input image and 10 types of $10 \times 10$ filters in a convolutional layer. The stride length in the convolution processing was set as $1 \times 2$. Fig. 5 shows the 10 types of filters used in this study. The following rectified linear unit (ReLU) function expressed by $r(x)$ was used as the activation function:

$$r(x) = \begin{cases} x & (x \geq 0) \\ 0 & (x < 0) \end{cases} \quad (4)$$

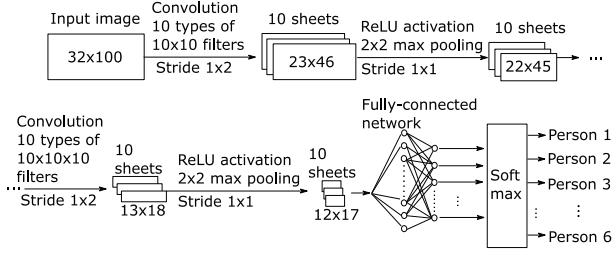

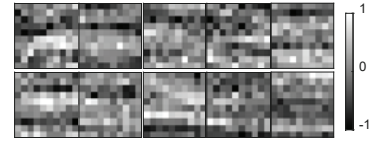

Fig. 5 Ten types of 10×10 filters used in the first layer.

Fig. 4 Double-layer CNN used for personal identification in this study.

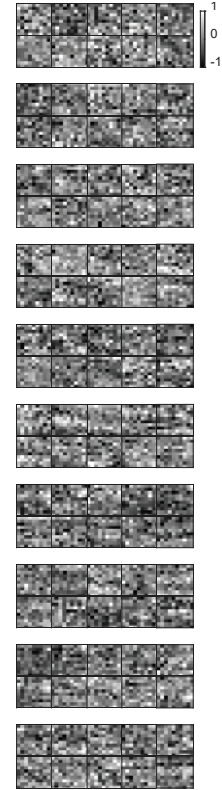

Fig. 6 Ten types of 10×10×10 filters used in the second layer. Type 1,2, ⋯,10 for each block from the top.

Next, max-pooling was used with pooling size and stride length of 2×2 and 1×1, respectively. Ten $22 \times 45$-pixel images were outputted after the convolution of the first layer, activation, and pooling processing. Next, the same processing was conducted again in the second layer. However, the output of the first layer in the second layer was regarded as 3D data with size $10 \times 22 \times 45$, and a 3D convolution with 10 types of $10 \times 10 \times 10$-pixel filters was conducted, as shown in Fig. 6. Ten $12 \times 17$-pixel images were outputted from the second layer, and the total number of pixels was $N_{\text{in}} = 2{,}040$. A fully connected layer was used with this information to obtain an output that corresponds to the number of subjects $N_{\text{out}}$. The $N_{\text{out}}$ values outputted from the fully connected layer between the input port number $N_{\text{in}}$ and the output port number $N_{\text{out}}$ were output as $q_1, q_2, \cdots, q_{N_{\text{out}}}$ after soft-max processing. Based on the number corresponding to the maximum output,

$$ i_{\text{est}} = \arg \max_i q_i \qquad (5) $$

a subject index $i_{\text{est}} \in \{1, 2, \cdots, N_{\text{out}}\}$ was obtained as the final identification result. This study involved six subjects, that is, $N_{\text{out}} = 6$, and the subject index $1, 2, \cdots, 6$ corresponds to the subject labels A, B, $\cdots$, F.

The spectrogram images obtained from the measurement data were classified into $N_{\text{out}}$ classes with subject labels attached. If the number of images in each class was set as $N_s$, the total number of images was $N_s N_{\text{out}}$. Of these, $N_l N_{\text{out}}$ images, where $N_l$ images were randomly selected from each class, were used for learning and the remaining $N_t N_{\text{out}} = N_s N_{\text{out}} - N_l N_{\text{out}}$ images were used for performance evaluation. Here, $N_t = N_s/k$, and cross-validation with k divisions was used. Specifically, $k = 10$, $N_s = 100$, $N_l = 90$, and $N_t = 10$.

The network was trained by optimizing the weights $w_{i,j}$ ($i = 1, \cdots, N_{\text{in}}, j = 1, \cdots, N_{\text{out}}$) of the fully connected layer. This corresponds to a $2{,}040 \times 6 = 12{,}240$ -dimensional nonlinear optimization problem in the application example considered in this study. The stochastic gradient descent method with momentum, which is widely used in CNN learning, was used for this optimization. The initial learning rate in this case was set as 0.001.

## IV. EVALUATION OF PERSONAL IDENTIFICATION ACCURACY

This section investigates the accuracy of personal identification using measurement data of walking and sitting motions based on the CNN discussed in the previous section. The $0 \le t \le 8$ s and $-3.5$ Hz $\le f_v < 3.5$ Hz sections in the spectrograms of the measurement data were converted into images. The image size of the original spectrogram was $32 \times 1600$ pixels; however, the number of pixels was reduced by thinning to produce a $32 \times 100$-pixel 8-bit grayscale image. The generated image was input into the CNN and used for network learning and accuracy evaluation. Supervised learning was used in this study; therefore, the actual subject labels were given as the correct answers. A total of 600 images were obtained because the walking and sitting motions and radar measurements were repeated 100 times for each of the six subjects. A 10-fold cross-validation was used for the performance evaluation. Of the 600 images, 540 randomly selected images were used for learning and the remaining 60 were used for performance evaluation.

Fig. 7 shows some of the images used in this study. These images are shown upside down relative to those in Fig. 3. The letters A–F in the figure refer to the subject labels. Of the 100 images obtained in the measurements of each subject, 10 are shown as examples. To train the CNN, many measurements were required; therefore, the maximum number of measurements per day was limited to five, with measurements performed intermittently over ~1 month. The spectrogram in Fig. 7 indicates that the patterns within an image varied for each measurement even in the same subject. Furthermore, the images showed clear differences in the spectrograms between subjects but also showed different patterns for the same individual. Therefore, personal identification based on manual feature extraction is difficult. To overcome this problem, automatic feature extraction using machine learning is required.

TABLE III
CONFUSION MATRIX (%) OF PERSONAL IDENTIFICATION

|  |  | Estimated Person | | | | | |
|---|---|---|---|---|---|---|---|
|  |  | A | B | C | D | E | F |
| Actual Person | A | 96.2 | 0.6 | 0.0 | 1.8 | 0.1 | 1.2 |
|  | B | 0.6 | 89.2 | 3.2 | 1.8 | 2.6 | 2.3 |
|  | C | 1.3 | 2.1 | 94.1 | 1.2 | 1.3 | 0.0 |
|  | D | 1.1 | 1.8 | 1.4 | 93.0 | 1.7 | 0.9 |
|  | E | 1.0 | 4.3 | 0.1 | 0.2 | 93.7 | 0.5 |
|  | F | 1.3 | 1.3 | 1.7 | 0.3 | 0.7 | 93.7 |

The weights of the fully connected layer prior to optimization were initialized with random numbers. Random numbers were also used for the stochastic gradient descent method with momentum used for optimization. Therefore, the personal identification performance changed depending on the initial values of the random numbers. To conduct performance evaluations in consideration of this effect, the initial values of pseudo-random numbers were changed to various values and the average performance was calculated and evaluated. Below, the initial value was changed in 100 ways. The 8-s spectrogram images, including both walking and sitting motions, were learned using a double-layer CNN, and the personal identification performance was evaluated. The learning epoch number for each trial was empirically set to 300. The identification accuracy was evaluated by inputting $N_t$ images into the network that had completed learning in each trial for accuracy validation. Different sets of images were used for accuracy validation and learning.

Table 3 shows the personal identification results obtained from the experiments as a confusion matrix. The row and column labels respectively indicate actual and estimated subjects. The diagonal elements of the confusion matrix correspond to the correct personal identification percentage. The average accuracy rate was 93.3%. The unbiased standard deviation of the accuracy rate calculated from 100 samples was 2.7%, and the two-tailed t value at a 95% confidence level was 1.99 when the parameter was 100; therefore, the 95% confidence interval ranged from 92.7% to 93.8%. This either exceeds or is almost equal to the accuracy rates of 90.9% [26], 78.4% [27], 80% [28], and 94.4% [29] achieved using conventional techniques. The average precision rate and reproducibility were both 93.6%. The F value, which is the harmonic mean of these precision and reproducibility values, was also 93.6%. The confusion matrix indicates that the accuracy rate of subject B was relatively low at 89.2%. The probability that the spectrogram of subject B was mistakenly estimated as that of subjects C, E, or F was relatively high at over 2%. Fig. 7 shows that the spectrograms of subjects B and D had different morphologies with each measurement and had relatively low reproducibility; this may explain the reduced average accuracy for these subjects.

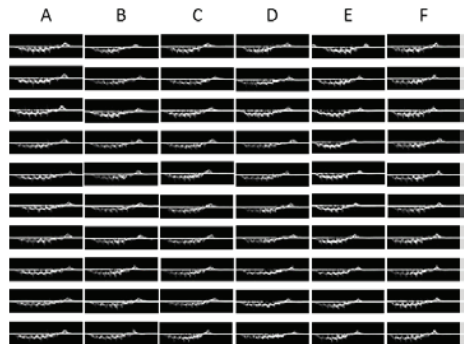

Fig. 7 Example of each of the 10 spectrogram images of walking and sitting motions of six subjects (A–F).

## V. DISCUSSIONS

### A. Types of Machine Learning

This study used a common double-layer CNN. Various methods can potentially be used for personal identification, including least-squares classification [33], support vector machine [34], KNN [35], decision trees [36], random forest, and the naïve Bayes method. Further, various types of neural networks are available, including those that use deep learning with multiple layers [37] or high-dimensional neural networks that use complex numbers or quaternions [38-40]. The personal identification method proposed in this study is only one example of a machine learning application.

### B. Input Image Size

Neural-network-based personal identification performance depends on various factors including the structure of and number of layers in the network and the amount of learning data. Further, the fully connected layer contains a large amount of data. Therefore, minimizing the dimensions of the input data is advantageous for reducing the learning time or avoiding overfitting. Accordingly, it is advantageous to keep the input image size small. This study used 32×100-pixel images as inputs; by contrast, Cao et al. [27] used 256×256-pixel images, Vandersmissen et al. [28] used 45×256-pixel images, and Yang et al. [30] used 100×100-pixel images. Reducing the image size requires consideration of the measurement time, sampling frequency, and the temporal and frequency resolutions that are determined by the maximum acceleration of the target as well as thinning to a range that satisfies the Nyquist condition. Simultaneously, the relationship between the degrees of freedom and the amount of learning data in the fully connected layer of the neural network needs to be considered.

In this study, the input image size was determined as follows. If the measurement time was set as $T_o$ and the sampling frequency was set as $f_s$, then the number of samples in the time direction was $M_s = T_o f_s$. The radar used in this study had a sampling frequency of $f_s = 200$ Hz, and $T_o = 8.0$ s was set in consideration of the length of time in the series of motions from walking to sitting, $M_s = 1,600$. The Nyquist velocity determined by the sampling frequency $f_s = 200$ Hz was 3.6 m/s. As shown in Fig. 3, a Doppler velocity that exceeded the Nyquist velocity during walking and sitting did not occur frequently, and therefore, there was no need to set the sampling frequency to a value higher than 200 Hz. The frequency resolution was approximately $\Delta f = 2/t_0$ for a window function $w(t)$ with effective width $t_0/2$ used in STFT, and the Doppler velocity resolution was $\Delta v = c/f_0 t_0$ when the central frequency was $f_0 = 4.2$ GHz. For example, $\Delta v = 0.45$ m/s if $t_0 = 0.16$ s. If the maximum acceleration of the human body was assumed to be 3.0 m/s$^2$, the maximum change in velocity during $t_0 = 0.16$ s would be 0.48 m/s. This value is similar to $\Delta v = 0.45$ m/s; therefore, there is no need for further extending $t_0$. Consequently, the number of samples in the Doppler velocity direction was determined to be $f_s t_0 = 32$ points. If a sampling interval $t_0/2$ where the window of the STFT overlaps by only half was selected to avoid oversampling in the time direction, then the number of samples in the time direction would be determined as $T_o/(t_0/2) = 100$ points. In this manner, the input image size for the CNN in this study was obtained as $32 \times 100$.

### C. Data Size for Training

Yang et al. [30] achieved a higher identification accuracy rate than that achieved in this study, but only by using a much larger number of images (22,500 vs 600) for learning. Generally, the performance improves with a larger amount of learning data. However, for the applications assumed in this study, it is unrealistic to have users perform the same movements repeatedly. Therefore, the actual amount of learning data is limited. In such cases, the effective amount of learning data can be increased by adding noise to the learning data or by performing contrast conversion in the data augmentation method. The introduction of auto-encoder-based pre-learning, transfer learning, and generative adversarial networks (GAN) may also result in improved performance. The introduction and investigation of these methods is an important topic for future study.

## VI. Conclusions

Personal identification technology will become increasingly important in the future in places shared by multiple people (e.g., home, workplace). In this regard, radar-based personal identification techniques are particularly important because they raise fewer privacy concerns. Previous studies have performed micro-Doppler measurements of walking human bodies to realize personal identification by using biological information such as breathing or heart rate. By contrast, this study uses human bodies sitting in a chair as subjects, measures walking and sitting motions by using ultrawideband radar, and performs personal identification by using micro-Doppler information obtained by time-frequency analysis. The subject motion was measured with an ultrawideband radar, converted to spectrograms using time-frequency analysis, and then further converted to a low-resolution grayscale image. Then, double-layer CNN-based identification was conducted. An average personal identification accuracy rate of 93.3% was achieved in experiments with six participants.

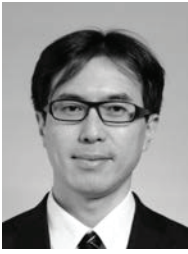 **Takuya Sakamoto** received the B.E. degree in Electrical and Electronic Engineering from Kyoto University, Kyoto, Japan, in 2000, and M.I. and Ph.D. degrees in Communications and Computer Engineering from the Graduate School of Informatics, Kyoto University, in 2002 and 2005, respectively.

From 2006 to 2015, he was an Assistant Professor at the Graduate School of Informatics, Kyoto University. From 2011 to 2013, he was also a Visiting Researcher at Delft University of Technology, Delft, the Netherlands. From 2015 to 2018, he was an Associate Professor at the Graduate School of Engineering, University of Hyogo, Himeji, Japan. In 2017, he was also a Visiting Scholar at the University of Hawaii at Manoa, Honolulu, HI, USA. From 2018, he has been a PRESTO Researcher at the Japan Science and Technology Agency, Kawaguchi, Japan. Currently, he is an Associate Professor at the Graduate School of Engineering, Kyoto University. His current research interests are system theory, inverse problems, radar signal processing, radar imaging, and wireless sensing of vital signs.

Dr. Sakamoto was a recipient of the Best Paper Award from the International Symposium on Antennas and Propagation (ISAP) in 2012 and the Masao Horiba Award in 2016. In 2017, he was invited as a semi-plenary speaker to the European Conference on Antennas and Propagation (EuCAP) in Paris, France.